# NATIONAL ACADEMIC DEPOSITORY: A STEP TOWARDS DIGITAL INDIA VISION

## SATINDER BAL GUPTA[a] AND MONIKA GUPTA[b1]

[a]Department of CSA, Indira Gandhi University, Meerup, Rewari, India
[b]Department of Chemistry, Vaish College, Rohtak, India

## ABSTRACT

The National Academic Depository of India is a distinctive, novel and progressive step visualized by Ministry of Human Resources Development, Govt. of India towards maintaining a database to hold the academic awards issued by Educational Institutions in an electronic and digital form. NAD promises to abolish the difficulties / inefficiencies of collecting, maintaining, and presenting physical paper certificates that can be easily copied / created and the verification processes which are costly, time consuming and disorganized. The depository can eradicate the need to store academic awards in physical form. It can verify the awards issued by different Institutions to the students in an easy way. The secure digital depository is a good proposal to do away with fake and forged certificates. The concept of academic depository is identical to the concept of financial securities. The pilot project is successfully completed with the help of Central Board of Secondary Education and some universities. In order to become fully functional, the depository has to conquer a few challenges with respect to academic diversities in terms of duration of courses and equivalence. National Academic Depository is a revolutionary effort towards the vision of Digital India.

Educational institutions issue academic certificates to students for their qualifications and the course of study at the institution. Presently, the database pertaining to the awards and related details are maintained by the educational institutions in physical form. There are many difficulties / inefficiencies linked with the existing method of collecting, maintaining, and presenting physical paper certificates. The most common of these are (LOKSABHA, 2011, NSDL, 2012)

- It is very difficult to save the printed/written records.
- It is very time consuming to retrieve the records from the registers because of accumulation of the data over the years.
- The verification procedure of these records is also very burdensome and time consuming.
- It is very difficult to obtain the duplicate mark sheets and degrees.
- It is not easy to obtain transcripts of the educational qualification.
- The bogus / forged awards can be created.
- The survival of many non-registered institutions and courses are also the biggest challenge.

To deal with all these limitations and issues of the Indian education system, the Ministry of Human Resource Development (MHRD) has initiated the process for establishment of a "National Academy Depository (NAD)" i.e. national database of academic qualifications in an electronic format. This database would assist online verification of academic awards issued by boards and universities and hence eliminate the chances to use bogus/ forged academic certificates to a great extent.

The NAD will provide interface to various agencies who want to authenticate certificate in online way whether such an academic certificate has been issued by the institution or not.

The MHRD ask over Central Board of Secondary Education to carry out pilot project so as to remove initial difficulties, if any. The pilot project has been undertaken with two agencies namely Central Depository Services (India) Ltd (CDSL) and National Securities Depositories Limited (NSDL) as these two are the only depositories registered with Security Exchange Board of India.

The pilot project was awarded to CDSL and NSDL in the month of Jan.2011. The agencies worked on the pilot project in close alliance with the CBSE and completed the pilot which has been built, owned and operated by these organizations. As a part of pilot implementation, CBSE has made available the academic awards of following examinations on the NAD portals (CBSE, 2011, CBSE2012).

- Central Teacher Eligibility Test (CTET) 2011
- CBSE Board XII Standard Exam 2011

This paper will give an overview of the functions, features, challenges and benefits of NAD to the various stakeholders.

[1]Corresponding author

## PROPOSED FUNCTIONS OF NAD

The NAD would perform the following functions (CBSE, 2011, CBSE2012):

i. The NAD will register various entities such as academic institutions granting admission for higher education, prospective employers, background checkers etc. in the depository.

ii. The NAD will provide the facility of uploading of details of academic qualifications by registered academic institutions such as boards or universities.

iii. The NAD will provide the facility of verification of the awards to the schools, colleges, and business firms etc. who are registered with NAD.

iv. The NAD will provide the facility of mapping of academic awards. This means the certificate holders who get registered with NAD have the facility of mapping their academic awards available in the system to their NAD account by submitting a request to NAD agent. Once the awards are mapped, certificate holder will be able to view his awards under a single view in his login.

**v.** The NAD will provide the facility to the stakeholders that they can request for issuance of 'Authentication Certificate'. Only registered certificate holder can request NAD to issue authentication certificate / mark sheet.

## THE CHALLENGES FOR THE NAD

The NAD has many challenges to overcome before it can provide this important and unique service. Some of the major challenges it has to overcome are as follows (Veera Gupta, 2013)

i. **Enormous number of accounts**: In general every student having education matric onwards should have an account in the depository. In the current situation, when basic and secondary education is made compulsory, the number of students graduating from the matric level is going to be more than ten million. In one side, a student has different roll numbers for different exams and on other side there are more than thirty boards and hundreds of universities providing the formal education.

ii. **Number of institutions in the country**: The knowledge Commission has recommended opening of three thousand universities as against existing four hundred. The vocational mission recommends opening of one lakh vocational centers in the country during next plan period. All the institutions eventually need to get registered with the depository.

iii. **Equivalence of Qualifications:** Till now, the qualifications offered by recognized school Boards and Universities to get jobs are having a mechanism to establish equivalence of degrees/awards. Currently, the education system is going through a change from annual exams to credit based system. The credits would help in standardization of academic units. So, the clarity on its framework is prerequisite for the depository.

Further, there are many training programs offered by the organizations for their employees. These are recognized by the employers for the purpose of employment but are out of the field of recognized qualification framework. The academic depository would create big challenge before the educational planners to decide on equivalence of awards offered by informal and in-service education providers.

iv. **Security of data**: The financial securities have faced IPO scam as the system lacked restrictive mechanism to check an individual from opening multiple accounts. The academic depository is also susceptible on account of determination of identity of the student.

## FEATURES OF NAD

The government of India is dedicated to bring administrative and academic reforms through the use of technology for providing efficient services to all stakeholders. The NAD would assist 24x7 online accesses to mark sheets, degrees etc. to educational institutions and employers with the permission of the candidates. The main features of NAD are as follows: (nad.ndml.in, ndml.in, skilloutlook.com, 2016, nad.co.in/NAD, groningendeclaration.org/signatories/cdsl-national-academic-depository)

- A system for Issuance, Hosting, Access of Digital Certificates and Online Verification.

- NAD will facilitate Academic Institutions to directly lodge the details of Academic Awards Issued in an online manner.

- The Academic Institution will include the details of the Aadhaar Number of the Student as part of the Certificate details so that Certificate can be securely made available to the concerned student.

- Certificate holders' will register on NAD system for accessing their certificate records. They will identify

themselves based on Aadhaar Verification framework provided by Unique Identification Authority of India.

- Registered students will have a unique account on NAD System and can access all their records 24/7 in an online secure manner.

- NAD can be accessed online by all authorised verifiers i.e. academic institutions admitting students for higher education, employers, banks, and government organizations for verification of academic awards of potential candidates. Verification users would not need to move to academic institutions for verification of the certificate records; they can directly approach NAD system and make a verification request for their certificate.

- NAD will report the concerned students about their verification request and if the students authenticate to share their certificate details, only then the record would be accessible to the verifying party for verification.

- NAD system has the essential security aspect to ensure that only certified users have access to authorized functions.

## BENEFITS OF NAD TO STUDENTS & CERTIFICATE HOLDERS

NAD provides academic certificates in digital format to students which are available for 24x7 online access to the students. The certificates are digital and verifiable and hence are easily trusted by all users. NAD is an extensively valuable proposal for students. Some most important benefits of NAD to students are as follows: (nad.ndml.in, ndml.in, skilloutlook.com, 2016, nad.co.in/NAD, groningendeclaration.org/signatories/cdsl-national-academic-depository)

- The online certificates will be available early as compared to physical certificates.

- The certificates will be online, 24x7 access to the records and no risk of losing and tearing etc.

- The online verifiable certificates are easily acceptable without requiring attested copies, original presentation.

- NAD has the facility for applying and obtaining duplicate copy of the certificates online very easily.

- NAD provides facility to submit verifiable copy of the certificate to employers, higher educational institutions.

## BENEFITS OF NAD TO ACADEMIC INSTITUTES

NAD integrates directly with Boards / Universities which issue Certificates and hence ensures authenticity of certificate records. University / Boards will directly keep the academic awards in NAD system in a safe online process. The records will identify the details of the students to whom such records belong by indicating the UID/Aadhaar number of the certificate holder. Major benefits of NAD to academic institutes are as follows: (nad.ndml.in, ndml.in, skilloutlook.com, 2016, nad.co.in/NAD, groningendeclaration.org/signatories/cdsl-national-academic-depository)

- NAD provides a digital front-end for academic institutions to provide online services for academic certificates.

- It decreases the cost and efforts for certificate issuance and verification activities.

- NAD provides digital back-end for maintaining up to date data of certificates.

- It reduces the threat of making fake & forged certificates.

- NAD provides reports & analysis about education.

## BENEFITS OF NAD TO CERTIFICATE VERIFICATION USERS

Through the depository, the employer can easily verify documents or certificates online and this can replace the existing process of manual verification by universities. Some major benefits to certificate verification users are as follows (nad.ndml.in, ndml.in, skilloutlook.com, 2016, nad.co.in/NAD, groningendeclaration.org/signatories/cdsl-national-academic-depository)

- NAD provides central system making verification of academic awards of different registered academic institutions at a single place.

- NAD offers online verification of academic awards and hence decreases the time and cost involved in the verification process.

- The processing of primary job application, loan application and admission application becomes fast and easy.

- As the system is fully online, clear and auditable, it lessens the need for intermediation and related risks.

## RESULTS OF THE NAD PILOT PROJECT AND CURRENT SCENARIO

A pilot project was carried out by the CBSE with the help of NSDL and CDSL. Both the agencies implemented the project in Java and it is J2EEE compliant. File validation facility is developed in java nd C++. The JBoss was used for application server and PostGres as RDBMS. The complete software system was configured on systems that support Intel processors (Veera Gupta, 2013).

The data of class SSC-II and the Central Teacher Eligibility Test (CTET) conducted by the CBSE in 2011 were uploaded in the database of the project. The total number of records of SSC-II was 770042 and the CTET was 794079 (Veera Gupta, 2013). This data was made available to the users for three months to know the results of the project. Approximately, 800 people visited the web site of NAD, approximately 300 people made registrations and about 100 verifications were performed during these three months (Veera Gupta, 2013). This shows that the project is successful.

From session 2016-17, the government has made it compulsory for the Central universities and centrally funded institutes to join the NAD. These and many other institutes have already joined the NAD and more are joining to stop forgery in education system.

## CONCLUSION

A lot of challenges are faced by the students and educational institutions for authorization of NAD. The depository may go along with educational institutions to initiate with. Both students and educational institutions store the data from so many years; the depository becomes full proof solution. If National Academic Depository is established effectively, it would prove to be advantageous for storing, retrieving and verifying the degrees. It will enhance the quality of teaching and learning by decreasing the administrative workload on staff recruited in educational institutions. The educational institutions may take up many more important tasks such as analysis of quality of test items, preparation of performance on items and feedback report to teachers, students and parents. It would help in allocating the main purpose of evaluation by enhancing quality of teaching and learning to educational institutions at all levels.

## REFERENCE

CBSE; 2012; National Academic depository; CBSE; Press release; 24th May.

Central Board of Secondary Education, 2011, Circular no. 74/2011, Oct. 11, www.cbse.nic.in

Gupta V., 2013, A National Academic Depository, 1st Annual International Interdisciplinary Conference, AIIC, 24-26 April, Azores, Portugal, pp. 390-397.

http://skilloutlook.com/2016/09/09/national-academic-depository-educational-degrees-awards-will-stored-digital-depository-like-securities-depository/

http://www.groningendeclaration.org/signatories/cdsl-national-academic-depository

https://nad.ndml.in/

https://www.nad.co.in/NAD/

https://www.ndml.in/national-academic-depository.php

Loksabha, 2011; Bill no.42; National Academic Depository Bill.

NSDL; 2012; Report on Proof of Concept of National Academic Depository Submitted to Central Board of Secondary Education (CBSE).